\documentclass[prl,showpacs,preprintnumbers,amsfonts,amssymb,amsmath,floats,twocolumn,superscriptaddress,aps]{revtex4}

\usepackage{graphicx}
\usepackage{dcolumn}
\usepackage{bm}
\usepackage[final,dvips]{epsfig}
\usepackage{bm}
\usepackage{color}

\newcommand{\pipi}{\ensuremath{(\pi,\pi)}}
\newcommand{\pio}{\ensuremath{(\pi,0)}}
\newcommand{\pit}{\ensuremath{(\frac{\pi}{2},\frac{\pi}{2})}}
\newcommand{\vv}[1]{{\mathbf #1}}
\newcommand{\eref}[1]{Eq.~\ref{#1}}

\newcommand{\kk}{\mathbf k}
\newcommand{\qq}{\mathbf q}
\newcommand{\be}{\begin{equation}}
\newcommand{\ee}{\end{equation}}
\newcommand{\ba}{\begin{eqnarray}}
\newcommand{\ea}{\end{eqnarray}}
\newcommand{\ff}[1]{{\bm #1}}

\begin{document}

\title{ 
The superconducting gap in the Hubbard model
and the two gap energy scales in high-T$_c$ cuprates
} 

\author{M. Aichhorn}

\affiliation{
Institute for Theoretical Physics, University of
W\"urzburg, Am Hubland, 97074~W\"urzburg, Germany
}

\author{E. Arrigoni}

\affiliation{
Institute of Theoretical Physics and Computational
Physics, Graz University of Technology, Petersgasse 16, 8010
Graz, Austria
}

\author{Z.\,B. Huang}

\affiliation{
Department of Physics, Hubei University, Wuhan 430062, PRC}

\author{W. Hanke}

\email{hanke@physik.uni-wuerzburg.de}
\affiliation{
Institute for Theoretical Physics, University of
W\"urzburg, Am Hubland, 97074~W\"urzburg, Germany
}

\begin{abstract}
       Recent excperiments (ARPES, Raman) suggest the presence of two distinct 
       energy gaps in high-$T_c$ superconductors (HTSC), exhibiting different doping
       dependences. Results of a variational cluster approach to the superconducting
       state of the two-dimensional Hubbard model are presented which show that
       this model qualitatively describes this gap dichotomy: One gap (antinodal)
       increases with less doping, a behavior long considered as reflecting
       the general gap behavior of the HTSC. 
       On the other hand, the near-nodal gap does even
       slightly decrease with underdoping. An explanation of this unexpected behavior 
       is given which emphasizes the crucial role of spin fluctuations in the pairing
       mechanism.
\end{abstract} 

\pacs{
71.10.-w,71.10.Hf,74.20.-z,74.72.-h 
} 

\maketitle

The celebrated pseudogap 
of high-temperature superconductors (HTSC) is
widely believed to be intimately related to the microscopic pairing
mechanism~\cite{ti.st}.
 Early angle-resolved photoemission spectroscopy
(ARPES) demonstrated that both the pseudogap and the superconducting
gap below T$_c$ are consistent with the $d_{x^2 - y^2}$ symmetry~\cite{lo.sh.96}. In
addition, both gaps were found to have a more or less identical doping
dependence, increasing when the doping was 
reduced~\cite{lo.sh.96,di.yo.96}.
This strongly supported a picture, where the pseudogap is a precursor to
the $d_{x^2 - y^2}$-superconducting (SC) state, i.e. the pseudogap
smoothly evolves into the SC gap, when phase coherence of the pairs
develops below $T_c$~\cite{ue.lu.89,ra.du.89,em.ki.95}. 
Accordingly, there seemed to be only
one energy scale in the system, namely the magnitude of the gap in the
antinodal region around \pio. However, this picture has been
challenged severely by Raman~\cite{op.ne.00,ta.sa.06} 
and very recent ARPES investigations~\cite{ta.le.06,ha.yo.06,ko.ta.06u}:
The experiments done below T$_c$ suggest two distinct energy gaps
exhibiting different doping dependencies.
While the gap
near the antinodal $(\pi, 0)$-region was still
found to strongly increase with decreasing doping, the other gap,
identified from sharp (coherent) peaks near the nodal region, has a
rather different doping dependence, i.e. does not increase with less
doping. These surprising observations have been interpreted as a
two-gap scenario, in which there are two energy scales, one
corresponding to the near nodal and one to the antinodal region around
the Fermi surface (FS). Clearly, these recent findings pose also a
substantial challenge to the microscopic theory.

In this letter, 
we show that the two-dimensional (2D) Hubbard model
accounts, at least qualitatively, 
for the observed different doping
dependencies of the gaps near the nodal and near the antinodal
regions. 
We suggest (see below) that
this ``two gaps'' 
scenario can be naturally explained by 
the anisotropy of the $d$-wave coupling strength at the Fermi surface,
which is related to the peaked structure of the spin-mediated pairing
interaction $V_{\rm Pairing}$ around $\vv Q_{AF}=\pipi$.
These results emphasize the importance of spin fluctuations in the pairing
mechanism for HTSC materials~\cite{c-spinfluct,be.sc.66,sc.lo.86,emer.86,mi.mo.90}. 
In addition, they support
the premise that the physics of the intriguing
HTSC materials should be contained in the 2D one-band Hubbard model. This 
premise has recently also obtained support from a variety
of cluster calculations which reproduce the overall ground-state
phase diagram of the HTSC as well as their single-particle
excitation spectra~\cite{ma.ja.05,se.la.05,ai.ar.06,ma.ja.06.2,ky.ka.06}.

Our main results, obtained by means of 
the Variational Cluster Approach (VCA)~\cite{po.ai.03,da.ai.03}, are
shown in Figs.~\ref{fig2} and \ref{fig5}. In  Fig.~\ref{fig2},
we plot the SC gap near the antinodal and near the nodal point as a function of
doping for the two-dimensional Hubbard model, whereby the 
 $d$-wave factor has been divided off for a better comparison.
The two gaps clearly behave differently as a function of doping, in
agreement 
with the experimental findings~\cite{ta.le.06,ha.yo.06,ko.ta.06u}. 
While the antinodal gap increases, the
nodal gap decreases with less doping.
However, an analysis of our results in terms of the 
anomalous self-energy 
indicates that below $T_c$ 
 there is only {\em one } gap mechanism. Both gaps are due
 to superconductivity, and
there is no contribution 
from the normal self-energy, i.e. a remnant of the normal-state
pseudogap.

As we will show below, the so-called ``two-gap'' behavior
 can be explained by the doping dependence of 
the spin-fluctuation mediated pairing interaction
$V_{\rm Pairing}(\vv q)$
which displays a strongly peaked structure around $\pipi$.
The width $a$ of this structure is given by the inverse antiferromagnetic
correlation length and, despite the reduction effect of 
vertex corrections, $a$ decreases with decreasing
doping~\cite{hu.ha.05,hu.ha.06}.
Due to the $d$-wave factor, the contribution to the superconducting gap
$\Delta_{SC}(\vv k)$
has a different dependence on the transfer momentum
$\qq$ of the effective interaction
depending on whether $\vv k$ is close to the nodal or antinodal
region.
This fact, connected with the doping dependence of the peak structure of
$V_{\rm Pairing}$ around $\pipi$ accounts for both the 
qualitatively different doping dependencies
of the gaps near \pio{} or near \pit, 
as well as for the
deviation of the SC gap from the nearest-neighbor $d$-wave form 
$\cos k_x -\cos k_y$ as observed in
experiments~\cite{ta.le.06,me.no.99,ko.ta.06u,ha.yo.06}.

Our results are obtained on the basis of two 
numerical techniques, which are appropriate for the treatment of
strongly-correlated materials, namely,
the Variational Cluster Approach
(VCA)~\cite{po.ai.03,da.ai.03} 
and a quantum Monte Carlo (QMC) cluster solution~\cite{hu.ha.05,hu.ha.06}.
The VCA  is based on the
self-energy functional theory, which was proposed and applied 
by Potthoff
{\em et al.}~\cite{pott.03.epjb1,pott.03.epjb2,ai.ar.06.2}. 
It provides a variational scheme to use dynamical
information from an exactly solvable ``reference'' system (in the VCA
        an isolated cluster) to go to the infinite-sized lattice fermion
system at low temperatures and at $T=0$ , in particular. 
The
ground-state phase diagram of the 2D 
Hubbard model 
\begin{equation}\label{eq:hamiltonian}%
H = \sum_{ij,\sigma}t_{ij}c_{i\sigma}^\dagger
c_{j\sigma}^{\phantom{\dagger}} + U\sum_i n_{i\uparrow}n_{i\downarrow}, 
\end{equation}
where $t_{ij}$ denote hopping matrix elements, $n_{i \uparrow}$ the
density at site $i$ with spin ``$\uparrow$'' and $U$ the local Coulomb
repulsion was calculated within the VCA 
by S\'{e}n\'{e}chal {\em et al.}~\cite{se.la.05}, and by
our 
group~\cite{ai.ar.05,ai.ar.06,ai.ar.06.2}.
For the cluster size used in the VCA (up to 10 site
clusters solved by a Lanczos technique),
the $T=0$ phase diagram of the Hubbard model in Eq.~(\ref{eq:hamiltonian}),
with hopping
terms up to third-nearest neighbors, correctly reproduces salient
features of the HTSC, such as the AF and dSC ground states in doping
ranges, which are qualitatively in agreement with electron- and
hole-doped cuprates.

\begin{figure}[t]
  \centering
  \includegraphics[width=0.9\columnwidth]{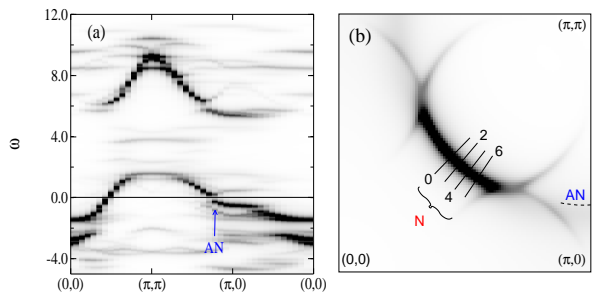}
  \caption{\label{fig1}%
    (Color online) Left (a): Spectral function of of the Hubbard model for 
    $U=8t$, $t^\prime=-0.3t$ and $x=0.07$ hole-doping, calculated with the
    VCA on a $3\times3$ cluster in the SC symmetry-broken phase. The blue arrow
    denotes the antinodal (AN) region. Right (b):
    The corresponding Fermi surface.
    The antinodal 
    region is marked by AN (blue), the nodal region by N (red). Solid lines indicate
    the momentum scans, 
    along which the gap is determined. Numbers refer to Fig.~\ref{fig2}.
  }
\end{figure}

Also the single-particle excitations agree with earlier
QMC data~\cite{pr.ha.97} and
with the main features observed in experiment~\cite{da.hu.04}: clear
signatures of the Mott gap, as well as the ``coherent'' quasiparticle
band of width $J$ ($\equiv$ magnetic exchange) and the
``incoherent'' lower and upper Hubbard bands are identified.

In Fig.~\ref{fig1}b 
we show a typical Fermi surface for the underdoped case at 
$x=0.07$ hole doping, obtained from the weight of the 
VCA single-particle spectral function at zero energy, $A({\vv k},\omega=0)$. 
The largest weight is found near the nodal direction,
marked by N in Fig.~\ref{fig1}, which corresponds to the ``Fermi
arc''. 
The weak features suggesting a splitting of the Fermi surface 
when  going to the antinodal region are  due to 
the finite cluster used as a reference system. Accordingly, we
measure the gap in the coherent
region around the node between line 0 and line 6
in Fig.~\ref{fig1}b, and at the antinodal region marked by AN 
in Fig.~\ref{fig1}.

\begin{figure}[t]
  \centering
  \includegraphics[width=0.7\columnwidth]{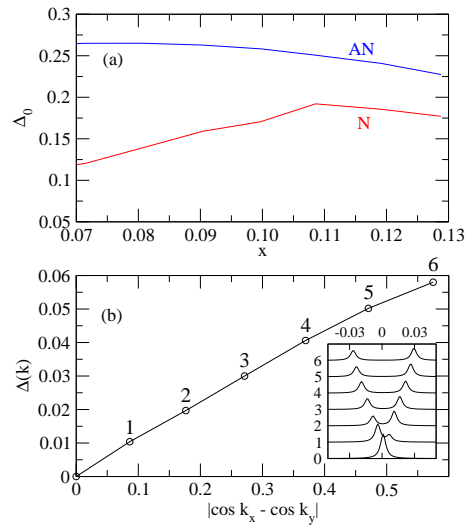}
  \caption{\label{fig2}%
    (Color online) (a) ``Normalized'' SC gap
    $\Delta_0\equiv 2 \Delta_{\vv k}/(\cos k_x -\cos k_y)$
    for $\vv k$ near the antinodal (dashed) and near the nodal FS point (solid)
    as a function of doping for the Hubbard model (parameters are given in
    the text).
    (b) SC gap as a function of the $d$-wave factor
    $(\cos k_x -\cos k_y)$ on the Fermi arc for $x=0.07$ hole doping. The
    numbers correspond to the momentum scans marked in Fig.~\ref{fig1}.
    Inset: 
    $A({\vv k},\omega)$ in a small window around the 
    FS for points 0 to 6.}
\end{figure}

In
Fig.~\ref{fig2}a we plot the (normalized) SC gaps
 $\Delta_0\equiv 2 \Delta_{\vv k}/(\cos k_x -\cos k_y)$
near the nodal and near the antinodal 
Fermi points as a function of  doping 
$x$.
Like in experiment, the amplitude of the d-wave gap
in the nodal region
{\em decreases} when driving the system from the optimal doped into 
the underdoped region, while it {\em increases} in the antinodal region.
The different doping dependences for the nodal and the antinodal regions
correspond to the
 two different energy scales observed in 
photoemission experiments. From the figure it is clear that
 $\Delta (\kk)$ cannot be described by a simple nearest-neighbor
d-wave form near both the nodal and the antinodal point, especially
for low dopings.
On the other hand, our plot of
$\Delta (\kk)$ displayed in Fig.~\ref{fig2}b
shows that in an extended region
around the nodal point  
the gap retains its
n.n. $d_{x^2-y^2}$-form, see also the inset of Fig.~\ref{fig2}b.
This latter finding is again in accordance
with experiment (see inset of Fig. 3 A in Ref.~\cite{ta.le.06}).

\begin{figure}[t]
  \centering
  \includegraphics[width=0.8\columnwidth]{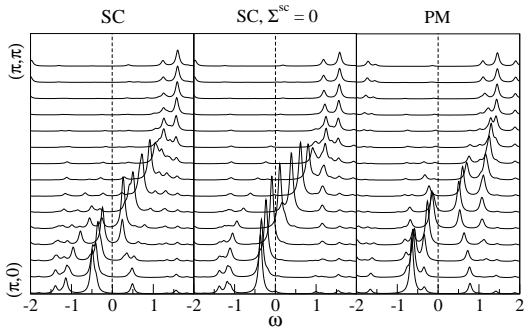}
  \caption{\label{fig5}%
    Spectral function from $(\pi,0)$ to $(\pi,\pi)$ at $h=0.07$ doping. 
    Left: $A(\ff k,\omega)$ in the SC broken phase. 
    Middle: SC phase, but $\Sigma^{\rm sc}=0$ (see text). Right: Paramagnetic solution.}
\end{figure}

For the understanding of this behavior it is first of all interesting 
to find out whether 
the gap originates only from one mechanism, related to superconductivity, or 
if there are 
competing 
mechanisms producing different gaps. 
In our calculation, 
we can address this question in the following way.
In matrix-Nambu formalism the Dyson equation reads
\begin{equation}
  G_{\alpha\beta}^{-1}(\ff Q,\omega) = \left(\begin{smallmatrix}
    \omega - T_{\alpha\beta}(\ff Q) - {\Sigma}_{\alpha\beta}^{\rm no}(\omega) & - {\Sigma}_{\alpha\beta}^{\rm sc}(\omega)\\
    - {\Sigma}_{\alpha\beta}^{\rm sc}(\omega) & \omega + T_{\alpha\beta}(\ff Q) + {\Sigma}_{\alpha\beta}^{\rm no^*}(-\omega)\end{smallmatrix}\right)\:,
\end{equation}
with $T_{\alpha\beta}(\ff Q)$ the hopping matrix, and $\alpha$, $\beta$ being the quantum
numbers within the reference system, i.e., the cluster sites.
After the variational opimization procedure, we 
set the SC-related parts of the self-energy $\Sigma_{\alpha\beta}^{\rm sc}$ 
to zero and evaluate  the corresponding spectral function. 
By comparing the left (full SC solution) and the 
middle panel ($\Sigma_{\alpha\beta}^{\rm sc}=0$) in Fig.~\ref{fig5}, 
one can clearly see that the gap
around $(\pi,0)$ is {\em only}{} due to $\Sigma_{\alpha\beta}^{\rm
  sc}$, and there are
no signatures of a remaining underlying pseudogap produced by a
mechanism different from superconductivity. 
Fig.~\ref{fig5} shows results for $x=0.07$ hole doping, but
we have checked that this behavior is the same in the whole doping range under investigation.

Since it is known that above $T_c$ the Hubbard model shows a pseudogap behavior \cite{pr.ha.97,ma.ja.06.2}, we also
compare our superconducting results with ``normal''-state solutions,
where we do not allow for $U(1)$ 
symmetry breaking in the 
variational 
procedure. The right panel of Fig.~\ref{fig5} shows that in this solution there is
indeed a pseudogap. Interestingly, this pseudogap increases with lowering the doping and, thus, shows a doping
dependence in qualitative agreement with finite-$T$ QMC simulations \cite{pr.ha.97} and also ARPES experiments, 
identifying the pseudogap above $T_c$ with the leading edge features
near $(\pi,0)$ \cite{lo.sh.96,di.yo.96,ha.yo.06}.
Our calculation is at $T=0$. However, taking 
all the above results together, they indicate that the breaking of the
$U(1)$ symmetry destroys the ``normal-state'' origin of the pseudogap,
and the gap ``below $T_c$'' is 
only due to superconductivity.

What is the physical reason for 
the different doping dependence of the nodal and antinodal gap ?
We  argue that this behavior is related to
the anisotropy of the $d$-wave coupling strength at the Fermi surface
in combination with the doping dependence of the peaked structure of the 
 spin-fluctuation mediated 
pairing interaction including full vertex corrections.
With a previously developed formalism~\cite{hu.ha.06} based on QMC, this effective 
pairing strength including the renormalization of 
the density of states can be calculated. 
\begin{figure}[t]
  \centering
  \includegraphics[width=0.7\columnwidth]{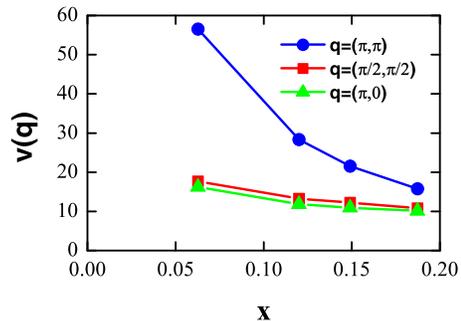}
  \caption{\label{figqmc}%
   (Color online) The doping dependence of $v(q)$ at $q=(\pi,\pi)$, $(\frac{\pi}{2},\frac{\pi}{2})$
   and $(\pi,0)$ for $U=8t$, calculated with QMC on an $8 \times 8$ cluster 
    for inverse 
     temperature $\beta=2.5t^{-1}$.
   Notice that the interaction hardly depends on the fermionic momentum $\vv k$, so only the
   dependence on $\vv q$ is shown for $\vv k=(-\pi,0)$.}
\end{figure} 
Results obtained for the $t$-$t^\prime$-$U$ (2D) Hubbard model 
showed that the effective pairing interaction 
$V_{\rm Pairing}=v(\kk,\qq)$
mediated by AF spin fluctuations 
is significantly influenced and reduced 
due to vertex corrections, 
but it clearly remains 
peaked around $\pipi$ \cite{hu.ha.05,hu.ha.06}.
This latter fact is, of course, well known from earlier unrenormalized
random-phase approximation calculations.
Most importantly, for transfer momenta $\vv q$ near $\pipi$, the
effective interaction (evaluated at the lowest Matsubara frequency) 
sharply increases with decreasing doping, see Fig.~\ref{figqmc}.
In contrast, for transfer momenta away from $(\pi,\pi)$, such as
$\qq=\pio$, $v(k,q)$ only shows at best a very weak increase with doping.

\begin{figure}[t]
  \centering
  \includegraphics[width=0.8\columnwidth]{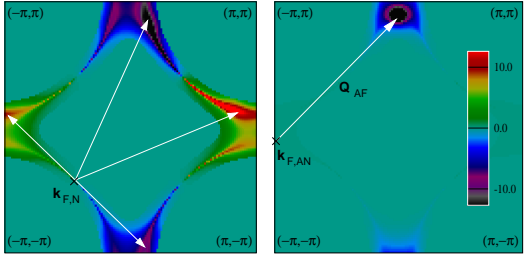}
  \caption{\label{figbcs}%
    Contributions to the gap $\Delta(\vv k_F)$ in \eref{bcseq} as function of
    $\vv k'$, divided by the d-wave factor $\cos(k_{F,x})-\cos(k_{F,y})$ for normalisation reasons. 
    Typical parameters have been used, $\mu=-1.0t, t'=-0.3t$.
    Left: Fermi momentum $\vv k_{F,N}$ near the nodal direction. Arrows indicate
    transfer vectors with large weights. Right: Fermi momentum $\vv k_{F,AN}$
    near the antinodal direction. The arrow is the AF momentum $\vv Q_{\rm AF}$.
    The color scale for both plots is given in the right plot.
  }
\end{figure}

Although HTSC materials  are clearly strongly correlated, our
argument can be understood qualitatively by means of a simple
BCS gap equation for
$\Delta(\vv k_F)$ at the Fermi point $\vv k_F$:
\be
\label{bcseq}
\Delta(\vv k_F) = - \int \frac{d^2 k'}{(2 \pi)^2} v(\vv k'-\vv k_F)
\frac{\Delta(\vv k')}{E(\vv k')} \;.
\ee
Here, $E(\vv k') = \sqrt{\xi_{\vv{k}'}^2 + \Delta(\vv k')^2}$, 
$\xi_{\vv k'}$ being the single-particle energy 
measured from the chemical potential.
After integrating over momenta perpendicular to the Fermi Surface in a
thin energy shell of width $\Omega \propto J\ll t$, one obtains for 
$\Delta \ll \Omega$:
\be
\label{bcseq2}
\Delta(\vv k_F) \approx - \int d \widehat{{\vv k}}_F' \ n(\vv k_F')  \ 
v(\vv k_F'-\vv k_F)\ \Delta(\vv k_F') \log 
\frac{\Omega}{\left| \Delta(\vv  k_F')\right|}
\;,
\ee
with
$n(\vv k_F')$ the density of states
per unit area of the FS at $\vv k_F'$.
The integral  in \eref{bcseq2} 
extends over the Fermi surface $d \widehat{\vv  k}_F'$, 
the $\log$ comes form the
integration of the  energy denominator, and we have exploited the fact
that $v$ depends mainly on the momentum transfer $\vv q=
\vv k_F'-\vv k_F$.
For $\vv k_F$ near the antinodal point $\pio$, the strongest
contribution
(normalized to the value of $v(\vv q)$) comes from 
$\vv k_F'$ near the two antinodal points
$(0,\pi)$ and $(0,-\pi)$, i. e. for
$\vv q=\vv k_F-\vv k_F'$ near $\pipi$. It is thus clear  from
the behavior of $v(\vv q)$  as a function of doping shown in
Fig.~\ref{figqmc} that one expects   
$\Delta(\vv k_F\sim \pio)$ to decrease fast with increasing doping.
On the other hand, for 
$\vv k_F$ around the nodal point $\pit$, the  contribution
for other nodal points cannot be strong since it is suppressed by the
nodes in $\Delta(\vv k_F')$. As a result, the strength of the SC gap is
controlled by  $v(\vv q)$ at momenta away from $\vv q=\pipi$. 
In order to eludicate this point, we plot 
in Fig.~\ref{figbcs} the contributions to the integral \eref{bcseq} 
for typical parameters $\mu=-1.0t$, $t'=-0.3t$, which show exactly the
behavior described above.
Hence, the behavior of  $v(\vv q)$ for $\vv q$ away from $\pipi$ shown in
Fig.~\ref{figqmc} provides the
physical
reason for which the SC gap at the nodal point does not increase
with decreasing doping \cite{ab.ch.01,bcs}.

Summarizing, we have shown that VCA calculations 
within the $t$-$t^\prime$-$U$ Hubbard model  
qualitatively reproduce the different doping dependences of the
superconducting gap recently seen in experiments. 
When going to lower hole dopings, the gap on the Fermi arc, near
the nodal region, {\em decreases}, whereas the gap near the 
antinode {\em increases}.
Moreover we have shown that there is no indication in our results for a
``two-gap'' scenario with distinct superconducting and pseudogap below $T_c$. Instead,
we found that obviously the SC gap in the symmetry broken solution absorbs the 
``normal-state'' pseudogap, resulting in a ``one-gap'' scenario with only a SC gap below $T_c$ 
exhibiting a more complicated structure in momentum space.
This more complicated structure 
is shown to be naturally 
explained in terms of
a spin-mediated pairing mechanism.
 
We thank 
Z.-X. Shen, A.\,V.~Chubukov and E.~Schachinger for constructive discussions, and in particular 
D.\,J.~Scalapino for his comments on the SC self-energy.
This work was supported by the Deutsche Forschungsgemeinschaft, grant
FOR 538
 and by the  Austrian science fund (FWF project P18551-N16).
  Z.B.H. was supported in part by NSFC Grant No.~10674043.


\begin{thebibliography}{10}
\bibitem{bcs}
We stress that this BCS calculation has been done in order to provide a first
  qualitative understanding of our theoretical results shown in
  Fig.~\ref{fig2}. We have checked that this behavior is also obtained within
  an Eliashberg calculation ~\cite{scha}, which is more appropriate for strong
  coupling and which takes into account the frequency dependence of the gap.
\end{thebibliography}
\end{document}